\documentclass[%
reprint,
superscriptaddress,
 amsmath,amssymb,
 aps,
 physrev,
]{revtex4-1}

\DeclareMathOperator{\csch}{csch}

\usepackage{graphicx}
\usepackage{dcolumn}
\usepackage{bm}
\usepackage{physics}
\usepackage{siunitx}

\usepackage{soul}

\usepackage{xr-hyper}
\usepackage{hyperref}
\hypersetup{
    colorlinks = true,
    allcolors = {blue},
    urlcolor = {black}
}
\usepackage[capitalize]{cleveref}
\usepackage{xfrac}
\usepackage{advdate}

\usepackage{mathtools}
\usepackage{xcolor}

\NewDocumentCommand{\evalat}{sO{\big}mm}{%
  \IfBooleanTF{#1}
   {\mleft. #3 \mright|_{#4}}
   {#3#2|_{#4}}%
}

\begin{document}

\title{Advantage of Coherent States in Ring Resonators over Any Quantum Probe Single-Pass Absorption Estimation Strategy}

\author{Alexandre Belsley}
\email{alex.belsley@bristol.ac.uk}
\affiliation{%
Quantum Engineering Technology Labs, H. H. Wills Physics
Laboratory and Department of Electrical \& Electronic Engineering, University of Bristol, BS8 1FD, United Kingdom
}%
\affiliation{%
Quantum Engineering Centre for Doctoral Training, H. H. Wills Physics Laboratory and Department of Electrical \& Electronic Engineering, University of Bristol, BS8 1FD, United Kingdom
}%
\author{Euan J. Allen}%
\affiliation{%
Quantum Engineering Technology Labs, H. H. Wills Physics
Laboratory and Department of Electrical \& Electronic Engineering, University of Bristol, BS8 1FD, United Kingdom
}%
\affiliation{%
Centre for Photonics and Photonic Materials, Department of Physics, University of Bath, Bath BA2 7AY, United Kingdom
}%
\author{Animesh Datta}
\affiliation{%
Department of Physics, University of Warwick, Coventry CV4 7AL, United Kingdom
}%
\author{Jonathan C. F. Matthews}
\email{jonathan.matthews@bristol.ac.uk}
\affiliation{%
Quantum Engineering Technology Labs, H. H. Wills Physics
Laboratory and Department of Electrical \& Electronic Engineering, University of Bristol, BS8 1FD, United Kingdom
}%

\date{\AdvanceDate[-1]\today}

\begin{abstract}
Quantum states of light have been shown to enhance precision in absorption estimation over classical strategies. By exploiting interference and resonant enhancement effects, we show that coherent-state probes in all-pass ring resonators can outperform any quantum probe single-pass strategy even when normalized by the mean input photon number. We also find that under optimal conditions coherent-state probes equal the performance of arbitrarily bright pure single-mode squeezed probes in all-pass ring resonators.
\end{abstract}
\maketitle

Quantum metrology seeks to determine and attain the fundamental quantum limits in estimating physical parameters~\cite{Polino2020Jun}. Its primary focus is to identify quantum strategies that outperform classical sensing schemes for an equivalent set of resources~\cite{Thomas-Peter2011Sep}. For example, given a mean number of probe photons, non-classical states have been used to enhance precision in estimating both phase and absorption in various applications including interferometry, magnetometry, and spectroscopy~\cite{Higgins2007Nov,Lang2014Aug,Slussarenko2017Nov,Moreau2017Jul,Lawrie2019Jun}.

Detecting and characterizing analytes using optical ring resonators has been applied in a wide-range of scenarios such as gas sensing~\cite{Hansel2020Jan}, measurements of mechanical strain~\cite{Campanella2013Dec}, and biochemical analysis~\cite{Luchansky2012Jan}.
The fundamental limits in estimating analyte properties using these structures is a largely unexplored topic in quantum metrology. A goal of this study is to quantify whether engineered photonic circuits with a classical light source can outperform non-classical state probes in a standard single-pass (SP) scheme. Compared to SP strategies, resonant optical cavities raise the prospect of enhanced precision both as a result of a buildup of the optical intensity and the increased number of interactions.

In this work, we quantify the magnitude of these precision gains when estimating the absorption coefficient and refractive index changes induced by an analyte evanescently coupled to an all-pass ring resonator. Using quantum estimation theory, we determine the experimental parameters that yield the highest possible precision for single-mode Gaussian probe states. At the optimal operating point, we find there is no advantage in using bright squeezed states over coherent state probes. Furthermore, coherent-state probes in all-pass ring resonator systems are capable of outperforming SP strategies with quantum probes, including squeezed light and Fock states with the same mean input photon number.

\begin{figure}[t]
    \centering
    \includegraphics[width=0.8\columnwidth]{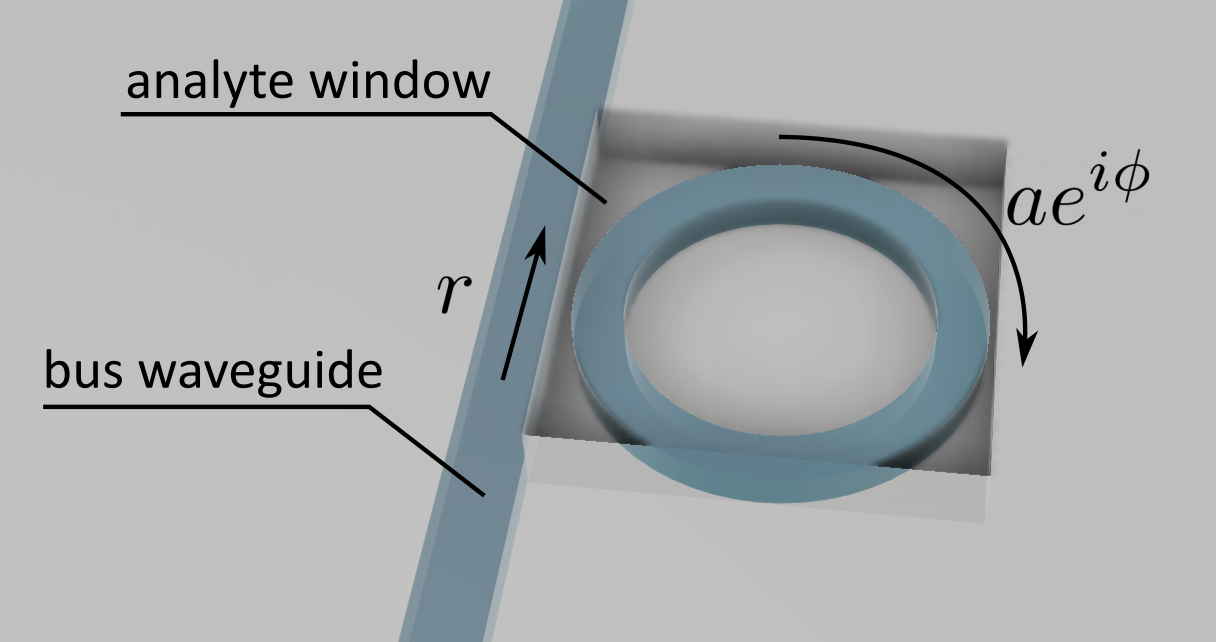}
    \caption{All-pass ring resonator with a self-coupling coefficient $r$, round trip phase $\phi$ and attenuation $a$. We seek to estimate the absorption coefficient $\alpha_A$ or refractive index $n_A$ of an analyte evanescently coupled to the ring resonator.}
    \label{fig:Ring_sketch}
\end{figure}

The system we consider is an all-pass ring resonator comprised of a ring resonator coupled to a bus waveguide, as depicted in  \cref{fig:Ring_sketch}. This contrasts to a SP strategy where the analyte is slotted into or surrounds a single-bus waveguide~\cite{Hansel2020Jan}. Light traveling in the ring is evanescently coupled to an analyte with an unknown absorption coefficient $\alpha_A$, which we seek to estimate.

The intensity transmission of this system is~\cite{Yariv2000Feb}
\begin{equation}
    \eta_R = \frac{a^2 - 2 r a \cos \phi + r^2}{1- 2 r a \cos \phi + (r a)^2}\,,
\label{Eq:allpass_transm_eta}
\end{equation}
where $r$ is the self-coupling coefficient, $\phi$ is the round-trip phase and $a$ is the attenuation coefficient.
\noindent A phase shift
\begin{equation}
    \theta_R = \pi + \phi  + \arctan \frac{r \sin \phi}{a - r \cos \phi} + \arctan \frac{r a \sin \phi}{1 - r a \cos \phi}\,,
\label{Eq:allpass_phase_theta}
\end{equation}
\noindent is imparted on the optical mode of the bus waveguide. The buildup factor, \textit{i.e.} the ratio between the circulating intensity in the ring and the incident intensity is
\begin{align}
B = \frac{{\left( {1 - {r^2}} \right){a^2}}}{{1 - 2ra\cos \phi  + {{\left( {ra} \right)}^2}}}\,.
\label{eq:buildup_factor}
\end{align}
\noindent Note that $B$ is independent of the intensity as the all-pass ring resonator is a linear system.

From the Beer-Lambert law, $a=e^{-\alpha_T L/2}$ where $L$ is the ring length and $\alpha_T$ the total linear absorption coefficient, which has two contributions
\begin{equation}
\alpha_T = \alpha_I + \Gamma \alpha_A\,. 
\end{equation}
Here $\alpha_I$ characterizes the intrinsic ring-waveguide loss in terms of an effective absorption coefficient and $\alpha_A$ is the targeted analyte absorption coefficient. The fraction of guided light in the ring that interacts with the analyte is quantified by the confinement factor $\Gamma$.

The all-pass ring resonator can be modeled as a channel $\Lambda$ that imparts a loss $\sqrt{1-\eta_R (\alpha_A)}$ and a phase shift $\theta_R (\alpha_A)$ on the probe state. 

\paragraph*{Fundamental quantum limit ---}The precision with which $\alpha_A$ can be estimated is bounded by~\cite{Helstrom1969Jun}
\begin{equation}
    \Delta^2 \alpha_A \geq \frac{1}{\nu\, \mathcal{Q}(\alpha_A)} \,.
\label{eq:QCRB_CRB}
\end{equation}
\noindent For a given experimental strategy repeated $\nu$ times, the Quantum Cram\'er-Rao bound (QCRB) relates the variance $\Delta^2 \alpha_A$ to the Quantum Fisher Information (QFI), $\mathcal{Q}(\alpha_A)$~\cite{Braunstein1994May}. It specifies the best precision achievable for a given channel and probe state.

\paragraph*{Coherent-state probes ---}We now quantify the performance of a coherent state probe in estimating $\alpha_A$. A coherent state $\ket{\beta}$ with mean photon number $\abs{\beta}^2$ is fully characterized by a displacement vector $\boldsymbol{d}$ with elements $d_i = \langle \hat{x}_i \rangle$ and a matrix $\boldsymbol{\Sigma} = \mathcal{I}/2$ of covariances of the quadrature operators $\hat{x}_1 = (\hat{a}^\dagger + \hat{a})$ and $\hat{x}_2 = i (\hat{a}^\dagger - \hat{a})$~\cite{Loudon1987Jun}. Application of the channel transformation $\ket{\beta}  \overset{\Lambda}{\mapsto} \ket{\sqrt{\eta_R}\, e^{i \theta_R} \beta}$ results in a displacement vector $\boldsymbol{d} = \beta \sqrt{\eta_R} \left(\cos\theta_R,\sin\theta_R\right)^{T}$.

The QFI of a single-mode Gaussian state is~\cite{Pinel2013Oct}
\begin{equation}
    \mathcal{Q}_G = \frac{\Tr[(\boldsymbol{\Sigma}^{-1} \boldsymbol{\Sigma}')^2]}{2 (1+P^2)} + \frac{2 {P'}^2}{1-P^4} +  {\boldsymbol{\Delta \text{X}}'}^\intercal\, \boldsymbol{\Sigma}^{-1}\,\boldsymbol{\Delta \text{X}}',
\label{Eq:QFI_Gaussian_State}    
\end{equation}
\noindent where $\bullet' \equiv \partial_{\alpha_A}\bullet$, $P = \det(\boldsymbol{\Sigma})^{-1/2}$ is the purity of the state and ${\boldsymbol{\Delta \text{X}}'} = \dd \langle {\boldsymbol{\text{X}}}_{\alpha_A+\epsilon} - {\boldsymbol{\text{X}}}_{\alpha_A}\rangle/ \dd \evalat{\epsilon}{\epsilon=0}$ with ${\boldsymbol{\text{X}}} = (\hat{x}_1,\hat{x}_2)$. For a coherent-state probe, the first two terms in \cref{Eq:QFI_Gaussian_State} are null, resulting in a QFI
\begin{equation}
\mathcal{Q}_C= \left( \abs{\beta} \, L\, \Gamma\,B\, e^{\alpha_T L/2}\right)^2\,.
\label{eq:QFI_coherent}
\end{equation}

\noindent as derived in Supplementary Material A \cite{Supp_Material_A}. \cref{eq:QFI_coherent} is maximized when $r = a$ and $\phi = 2\pi m, m \in \mathbb{Z}$, that is when the all-pass ring resonator is critically coupled and on resonance. Under these conditions, \cref{eq:QFI_coherent} reduces to
\begin{equation}
\mathcal{Q}_C \big\rvert_{r=a,\,\phi=2 \pi m} =\abs{\beta}^2\,\frac{L^2\,\Gamma^2\, B\,}{1- e^{-\,\alpha_T L }}\,.
\label{eq:QFI_coherent_critical}
\end{equation}

 \noindent Note that we assume that the incident optical field has a temporal coherence length that is sufficiently large to permit interference between light circulating in the ring and light entering the ring via the bus waveguide. In this respect, Fock state probes are beyond the scope of this work due to their wavepackets’ finite temporal coherence.

Instead of estimating loss, the above formalism readily allows one to estimate phase or equivalently refractive index changes $n_A$ induced by an analyte. This quantity is related to the round trip phase  $\phi = 2 \pi (n_I + \Gamma n_A)L/\lambda$, where $\lambda$ is the free-space wavelength and $n_I$ is the intrinsic refractive index. The only difference is a scaling factor $(4 \pi/\lambda)^2$, such that the QFI when estimating $n_A$ with a coherent-state probe under optimal operating conditions is
\begin{equation}
    \mathcal{Q}_C(n_A) = \abs{\beta}^2\, \left(\frac{4 \pi}{\lambda}\right)^2\, \frac{L^2\,\Gamma^2\, B\,}{1- e^{-\,\alpha_T L }}\,.
\label{eq:QFI_Quantum_Limit_nA}
\end{equation}
\paragraph*{Single-mode Gaussian probes ---}
We now determine the performance of pure single-mode Gaussian state probes. Using a numerical optimization algorithm, we find that the QFI in estimating $\alpha_A$ for these probes is also maximum at critical coupling and on resonance (see Supplementary Material A \cite{Supp_Material_A}). At this optimal operating point, the QFI is given by
\begin{equation}
    \mathcal{Q}_S =  \left(\abs{\beta}^2 + \sinh^2s\right)\frac{L^2\,\Gamma^2\, B\,}{1- e^{-\,\alpha_T L }}\,,
\label{eq:QFI_Gaussian}
\end{equation}
\noindent where $s$ is the squeezing factor.
Note that the term in parenthesis is equal to the mean number of photons in this bright squeezed state. Therefore, when normalized by the mean input photon number, an arbitrarily bright pure single-mode squeezed state performs at the same level as a coherent-state probe given in \cref{eq:QFI_coherent_critical}.

\begin{figure*}[t]
    \centering
    \includegraphics{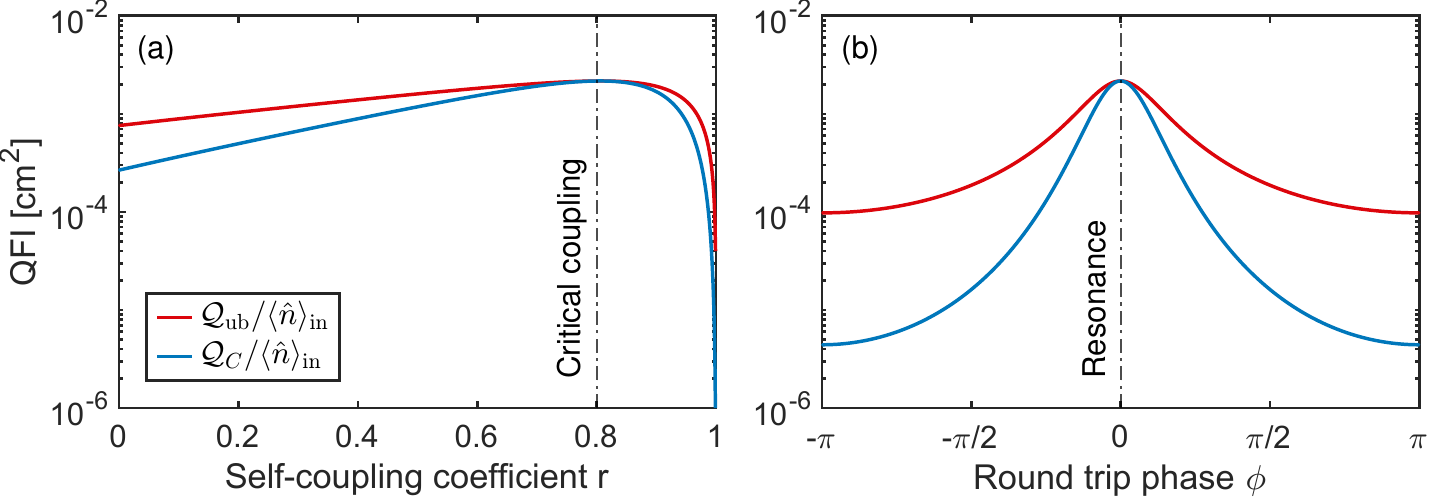}
    \caption{QFI when estimating the absorption coefficient $\alpha_A$ normalized by the mean input photon number $\langle \hat{n} \rangle_{\rm{in}}$ as a function of the all-pass ring resonator's (a) self-coupling coefficient $r$ and (b) round trip phase $\phi$. At critical coupling ($r = a = 0.8$) and on resonance ($\phi = 0$), the QFI for a coherent-state probe is maximum and equals the upper bound in \cref{eq:QFI_bound_ring_CPLE}, \textit{i.e}. $\mathcal{Q}_C = \mathcal{Q}_{\rm{ub}}$. Common parameters to both sub-figures are as follows: $\alpha_A = \SI{20}{\cm^{-1}}$, $\alpha_I = \SI{5}{\dB \per \cm}$, $\Gamma = 0.43$, and ring radius $R = \SI{75}{\micro\meter}$. $\phi=0$ was set in sub-figure (a) and $r=0.8$ in sub-figure (b).}
    \label{fig:QFI_Ring}
\end{figure*}

\paragraph*{Correlated phase and loss estimation ---}Birchall et al.~\cite{Birchall2020Apr} considered the situation where both the phase $\theta$ and loss $\eta$ imparted on a channel depend on a common parameter $\chi$. Using a unitary dilation with a single free environmental parameter, the following upper bound on the QFI was derived when estimating $\chi$ 
\begin{equation}
    \mathcal{Q}(\chi) \leq \langle \hat{n} \rangle_{\text{in}} \frac{4 \eta^2 (\partial_\chi \theta)^2 + (\partial_\chi \eta)^2}{\eta(1-\eta)} \,.
\label{eq:QFI_bound_CPLE}    
\end{equation}
\noindent Here $\langle \hat{n} \rangle_{\text{in}}$ is the mean number of input probe photons. For an all-pass ring resonator where the parameter of interest $\chi = \alpha_A$, this upper bound takes the form
\begin{equation}
     \mathcal{Q}(\alpha_A) \leq \langle \hat{n} \rangle_{\text{in}} \frac{L^2\,\Gamma^2\, B\,}{1- e^{-\,\alpha_T L }} \coloneqq \mathcal{Q}_{\rm{ub}} \,.
\label{eq:QFI_bound_ring_CPLE}    
\end{equation}
\noindent As shown in \cref{fig:QFI_Ring}, $\mathcal{Q}_{\rm{ub}}$ is identical to the QFI for a coherent-state probe at the optimal operating point given in \cref{eq:QFI_coherent_critical}. 
This upper bound is thus tight for pure single-mode Gaussian probe states.

\paragraph*{Optimal measurement strategy ---}Having derived the fundamental precision limit that can be achieved with pure single-mode Gaussian probes, we now show that intensity measurements are capable of saturating it.

The variance in the photon number $\hat{n}$ with this measurement strategy is
\begin{equation}
\langle \Delta^2 \hat{n} \rangle = {\eta_R}^2  \langle \Delta^2 \hat{n} \rangle_{\text{in}} + \eta_R (1-\eta_R) \langle \hat{n} \rangle_{\text{in}}\,.
\end{equation}
For a coherent-state probe, the input variance $\langle \Delta^2 \hat{n} \rangle_{\text{in}}$ and mean photon number $\langle \hat{n}\rangle_{\text{in}}$ are both equal to $\abs{\beta}^2$. The mean photon number at the output $\langle \hat{n} \rangle = \eta_R \abs{\beta}^2$. Using error propagation, the variance in estimating $\alpha_A$ for a resonant, critically-coupled ring resonator is
\begin{equation}
\Delta^2 \alpha_A = \langle \Delta^2 \hat{n} \rangle \abs{\frac{\partial \langle \hat{n} \rangle}{\partial \alpha_A}}^{-2} =\left(\abs{\beta}^2  \frac{ L^2\,\Gamma^2\,B}{1- e^{-\,\alpha_T L }}\right)^{-1},
\end{equation}
\noindent which is the reciprocal of \cref{eq:QFI_coherent_critical}. Thus, an intensity measurement saturates the QCRB. Note that under critical coupling and on resonance where the QFI is maximum, no light is transmitted. A null output in this case is not synonymous with no information. At critical coupling, $r=a$ and, provided one knows the all-pass ring parameters $\{\alpha_I,r,\Gamma,L\}$ with high precision, maximum information in estimating $\alpha_A$ is obtained. While operating at this optimal point can be experimentally challenging, slight detunings in $r$ and/or $\phi$ still yield near maximum QFI as shown in~\cref{fig:QFI_Ring}.

\paragraph*{Single-pass strategies ---}We now compare the performance of the all-pass ring resonator with SP strategies where the analyte is directly probed by an input quantum state. The transmission $\eta_{\text{SP}} = e^{- \alpha_A L}$ in the ideal case where propagation and reflection losses are neglected.

Fock states with photon number $N_0$ have been shown to be optimal in estimating the transmission of an analyte on a per input photon basis~\cite{Adesso2009Apr}. This is also the case when estimating $\alpha_A$, with an associated QFI~\cite{Allen2020Aug}
\begin{equation}
    \mathrm{QFI}_{F, \text{SP}} = N_0 \frac{ L^2}{e^{\alpha_A L}-1}\,.
\label{eq:QFI_SP_Fock}    
\end{equation}
\noindent This expression is maximized for an analyte length $L_{F,\text{opt}} \approx 1.59/\alpha_A$. Despite yielding the highest precision in estimating $\alpha_A$ for a fixed number of input photons in a SP strategy, Fock state probes are of limited practical use due to the difficulty in generating these states with high photon number.
Their performance can be readily surpassed by coherent-state probes with higher brightness as the QFI scales with the mean input photon number~\cite{Allen2020Aug}
\begin{equation}
    \mathrm{QFI}_{C, \text{SP}} = \abs{\beta}^2 L^2 e^{-\alpha_A L}\,.
\label{eq:QFI_SP_Coherent}
\end{equation}
\noindent The optimal analyte length in this case is $L_{C,\text{opt}} = 2/\alpha_A$.

Comparing \cref{eq:QFI_coherent_critical,eq:QFI_SP_Coherent}, we observe that a ring resonator amplifies the mean input  photon number by the buildup factor $B$ leading to an effective mean photon number $\abs{\beta}_{\text{eff}}^2 = B \abs{\beta}^2$ in the ring. Additionally, the effective analyte path length $L_{\text{eff}} = \Gamma L$ is decreased due to the finite coupling factor $\Gamma$ of the evanescent waves interacting with the analyte. Finally, the SP strategy transmission factor $e^{-\alpha_A L}$ is converted into $(1-e^{-\alpha_T L})^{-1}$ reflecting the fact that at critical coupling the light remains circulating in the ring until it is lost by an absorption or scattering event. Thus, in a critically coupled ring resonator, every photon interacts with the analyte in the limit of negligible intrinsic loss. The combination of the enhancement due to the buildup factor with high analyte interaction efficiency can result in an appropriately designed ring resonator structure reaching higher QFI values than SP strategies.

When intrinsic loss is present, not all of the input photons are absorbed by the analyte. Nevertheless, provided the intrinsic loss is small enough, the critically coupled all-pass ring resonator with a coherent-state probe will outperform an optimal SP strategy employing Fock state probes on a mean input photon number basis. The limiting value for the intrinsic loss under which this occurs can be readily obtained by comparing \cref{eq:QFI_coherent_critical,eq:QFI_SP_Fock} with $L_{F,\text{opt}}$, yielding the inequality
\begin{equation}
    \alpha_I L \alt 2 \csch^{-1}(1.61/L \Gamma \alpha_A) -\Gamma \alpha_A L\,.
\label{eq:ring_outperforms}    
\end{equation}

\begin{figure}[t]
    \centering
    \includegraphics{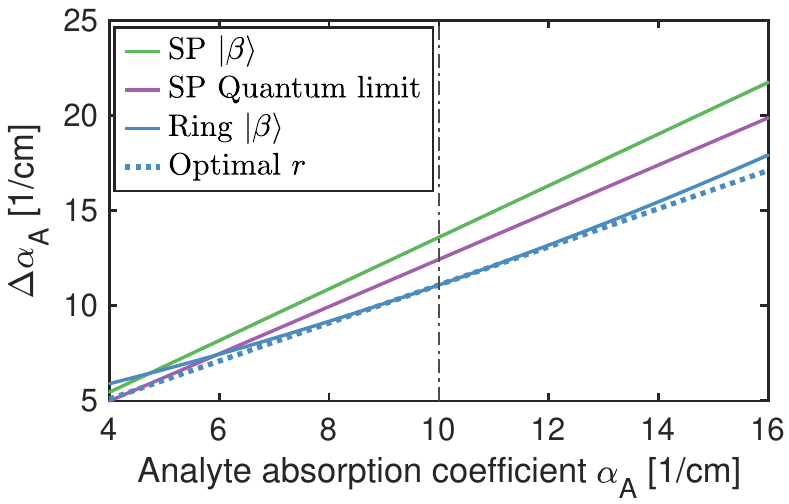}
    \caption{Standard deviations $\Delta \alpha_A$ normalized by the mean input photon number for an all-pass ring resonator probed by a coherent state (in blue), the quantum limit for a single-pass (SP) strategy attainable with a Fock state (in purple) and SP with a coherent-state probe (in green). For both SP strategies, the analyte length has been continuously optimized as $\alpha_A$ varies. The all-pass ring resonator is critically coupled for a target $\alpha_A = \SI{10}{\cm^{-1}}$. Optimizing the self-coupling coefficient $r$ further improves its performance (dotted blue). }
    \label{fig:Ring_vs_SP}
\end{figure}

As an example, we consider estimating the absorption coefficient of N-methylaniline near \SI{1500}{\nm} which has been previously studied using a classical approach by Nitkowski et al.~\cite{Nitkowski2008Aug}. The silicon all-pass ring resonator used had a radius $R = \SI{50}{\micro\meter}$ and confinement factor $\Gamma = 0.43$. Using a coherent-state probe, an average standard deviation $\Delta \alpha_A = \SI{0.25}{\cm^{-1}}$ for $\nu = 8$ trials  was reported. The peak absorption measured was approximately $\alpha_A = \SI{10}{\per\cm}$. With current technology, intrinsic silicon waveguide loss rates $\alpha_I < \SI{2.0}{\dB \per\cm} \approx \SI{0.6}{ \per\cm}$ are achievable~\cite{Bogaerts2012Jan}. In \cref{fig:Ring_vs_SP}, we plot the standard deviation $\Delta \alpha_A$ normalized by the mean number of probe photons for a ring resonator with the aforementioned parameters and a self-coupling coefficient $r = 0.93$ chosen to induce critical coupling at a target $\alpha_A = \SI{10}{\per\cm}$. This results in a standard deviation $\Delta \alpha_A = \SI{11.1}{\per\cm}$ per mean input photon number for a single trial. For comparison, we also plot the standard deviations achievable with Fock and coherent-state probes in ideal SP strategies where the length of the analyte has been continuously optimized as the absorption coefficient varies. At the target absorption coefficient $\alpha_A = \SI{10}{\per\cm}$, these are respectively 12\% and  23\% worse. Using \cref{eq:ring_outperforms}, the critically-coupled all-pass ring with optimal $r$ outperforms any SP strategy when $\alpha_A > \SI{4.4}{\per\cm}$ . For perspective, operating an all-pass ring resonator with $\alpha_I = \SI{2.0}{\dB \per\cm}$ at the optimal condition, a coherent-state probe with $\langle \hat{n} \rangle_\text{in} = 2000$ would be sufficient to reach a standard deviation better than that reported by Nitkowski et al.~\cite{Nitkowski2008Aug}.

Multi-pass (MP) strategies without resonant enhancement, where an analyte with a fixed length $L_0$ is traversed $k$ times by the incident light, have been proposed to improve precision beyond that of a SP strategy~\cite{Higgins2007Nov, Birchall2017Dec}. However, since for a given analyte thickness, the net effect of a MP strategy is to increase the analyte thickness by an integer multiple, it can never surpass the precision of a SP strategy with an optimal analyte length.

\begin{figure}[t]
    \centering
    \includegraphics[width=\columnwidth]{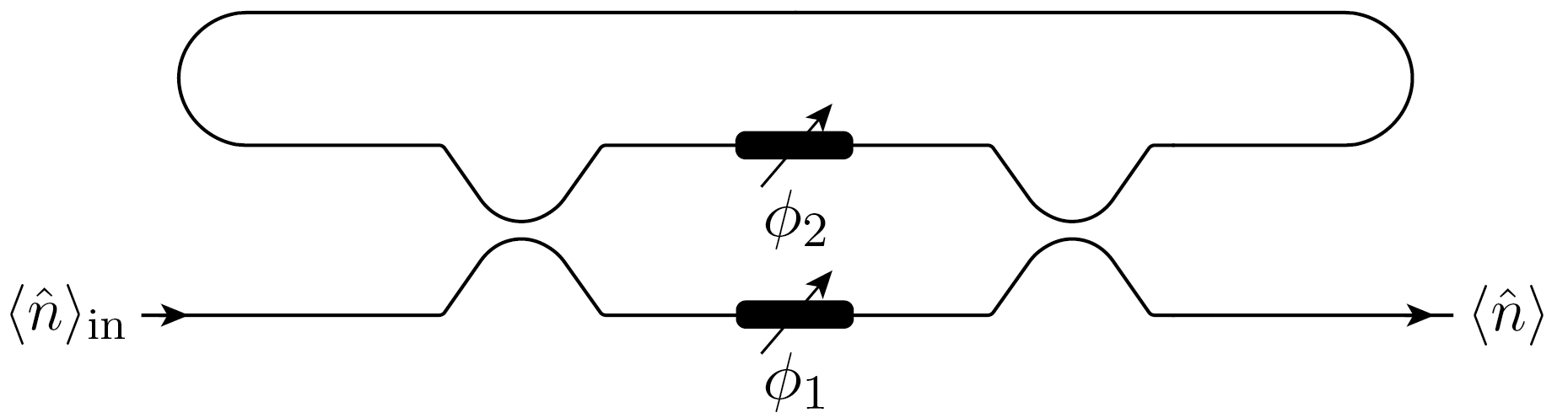}
    \caption{Mach–Zehnder interferometer-coupled ring resonator with tunable phase shifters $\phi_1$ and $\phi_2$.}
    \label{fig:Dual_Coupler_plot}
\end{figure}

\paragraph*{Tunable coupling regime ---}Operating at critical coupling maximizes the QFI in estimating $\alpha_A$ and $n_A$. This would require fabricating customized all-pass ring resonators for each individual analyte. This can be overcome by using a Mach–Zehnder interferometer-coupled ring resonator, which is formally equivalent to an all-pass ring with a tunable complex self-coupling coefficient~\cite{sacher2009coupler}
\begin{equation}
     \rho = i \exp[i (\phi_1 + \phi_2)/2] \cos[(\phi_1-\phi_2)/2],
\end{equation}
\noindent where $\phi_1$ and $\phi_2$ are the phases in the upper and bottom arms respectively (see \cref{fig:Dual_Coupler_plot}). By carefully tuning both of these phases, one can shift the operating point of the equivalent all-pass ring resonator from undercoupled to overcoupled passing through the desired critical coupling condition. For the case displayed in \cref{fig:Ring_vs_SP}, the all-pass ring resonator maintains its performance for analyte absorption coefficients within 20\% of the target value $\alpha_A = \SI{10}{\cm^{-1}}$, highlighting its robustness. 

\paragraph*{Conclusion ---}We have quantified the performance of pure single-mode Gaussian probes in estimating the absorption coefficient and refractive index changes induced by an analyte evanescently coupled to an all-pass ring resonator. We found the highest precision in estimating these parameters is achieved at critically coupling and on resonance. At this optimal operating point, there is no advantage in using bright single-mode squeezed states over coherent-state probes. Practically, operating at this ideal point can be facilitated by using a Mach–Zehnder interferometer-coupled ring resonator. 

There are many cases, such as when there is a limit in total optical probe power or competing noise sources in an experiment~\cite{LIGOScientificCollaborationandVirgoCollaboration2016Feb,Truong2011Mar}, where a key performance metric is the precision attainable by a given strategy normalized by the mean input photon number. On this basis, an all-pass ring resonator with coherent-state probes can yield higher precision than any single-pass strategy, including those using Fock or squeezed state probes with optimal analyte length.

Fully integrated, low-loss  ring resonator systems~\cite{Ji2017Jun} with shot-noise limited coherent-state sources~\cite{Fu2017Mar} and state-of-the-art detectors~\cite{Tasker2021Jan} can surpass the precision attainable with single-pass quantum probe sensors.
More generally, our results suggest that engineered photonic circuits are promising candidates for enhancing precision in parameter estimation. As is the case for the all-pass ring resonator, these precision gains can preclude the need for sophisticated quantum probe state generation and detection schemes.

Our findings are relevant for lab-on-chip resonator sensors~\cite{washburn2011photonics}, which have important practical applications ranging from environmental monitoring~\cite{Hodgkinson2012Nov} and ultrasonic imaging~\cite{Westerveld2021Mar} to antibody profiling~\cite{Shia2013Jan} and cancer detection~\cite{Shin2013Feb}.

\paragraph*{Acknowledgments ---}We thank Krishna C. Balram, Patrick M. Birchall, Osian Wolley, and Gabriel R. Higgins for helpful discussions. A.B. acknowledges support from the EPSRC grant EP/S023607/1. E.J.A. acknowledges support from an EPSRC doctoral prize from EP/R513179/1 and the Royal Academy of Engineering under the Research Fellowship scheme. J.C.F.M. acknowledges support from the EPSRC UK Quantum Technology Hub in Quantum Enhanced Imaging (QuantIC) EP/T00097X/1. A.B. and J.C.F.M. acknowledge support from the ERC starting grant ERC-2018-STG 803665. All data needed to evaluate the conclusions of the paper are present in the main text and the Supplementary Information.

\nocite{*}
\bibliography{mainbib.bib}

\clearpage
\onecolumngrid

\begin{center}
\textbf{\large SUPPLEMENTARY MATERIALS}
\end{center}

\setcounter{equation}{0}
\setcounter{figure}{0}
\setcounter{table}{0}
\setcounter{page}{1}
\makeatletter
\renewcommand{\theequation}{S\arabic{equation}}
\renewcommand{\thefigure}{S\arabic{figure}}
\renewcommand{\bibnumfmt}[1]{[#1]}
\renewcommand{\citenumfont}[1]{#1}

\section*{SUPPLEMENTARY MATERIAL A: Single-mode Gaussian probes}

Here we supplement the main text by quantifying the performance of pure single-mode Gaussian probe states in estimating the absorption coefficient of an analyte evanescently coupled to an all-pass ring resonator.

Consider an arbitrary pure single-mode Gaussian probe state $\ket{\psi_G} = \hat{R}(\varphi) \hat{D}(\beta) \hat{S}(\xi) \ket{0}$ with $\xi$ and $\beta$ arbitrary complex parameters. Here  $\hat{R}(\varphi) = \exp(i \hat{a}^\dagger \hat{a} \varphi)$, $\hat{S}(\xi) = \exp[\frac{1}{2} \xi^2 \hat{a}^{\dag 2} - \frac{1}{2} \xi^{* 2} \hat{a}^{2})]$ with $\xi = s e^{i \chi}$, and $\hat{D}(\beta) = \exp[\beta(\hat{a}^\dagger - \hat{a})]$ are the rotation, squeezing, and displacement operators respectively.

These states are fully characterized by a displacement vector $\boldsymbol{d}$ with elements $d_i = \langle \hat{x}_i \rangle$ and a matrix $\Sigma_{i,j} = \frac{1}{2} \langle \hat{x}_i \hat{x}_j + \hat{x}_j \hat{x}_i \rangle - \langle \hat{x}_i \rangle \langle \hat{x}_j \rangle$ of covariances of the quadrature operators $\hat{x}_1 = (\hat{a}^\dagger + \hat{a})$ and $\hat{x}_2 = i (\hat{a}^\dagger - \hat{a})$~\cite{Loudon1987Jun}. Applying the all-pass ring resonator channel transformation $\Lambda$, the covariance matrix becomes
\begin{equation}
    \boldsymbol{\tilde{\Sigma}} = \begin{pmatrix}
    1-2 \eta  \sinh s \cosh s \cos \Theta+2 \eta  \sinh ^2 s & 2 \eta  \sinh s \cosh s \sin \Theta\\ 2 \eta  \sinh s \cosh s \sin \Theta
    & 1+2 \eta  \sinh s \cosh s \cos \Theta + 2 \eta  \sinh ^2 s
    \end{pmatrix} \,,
\label{eq:cov_matrix_gaussian}  
\end{equation}

\noindent and the displacement vector $\boldsymbol{\tilde{d}} = 2 \abs{\beta} \sqrt{\eta}  \begin{pmatrix} \cos \Theta\\ \sin \Theta \end{pmatrix}$ where $\Theta = 2 \text{$\theta_R$}+2 \varphi -\chi$.

The QFI for this probe state is given by \cref{Eq:QFI_Gaussian_State}. The first term in this equation, that is the evolution of the noise properties of the state encoded in $\boldsymbol{\tilde{\Sigma}}$, is given by
\begin{align}
    \frac{\Tr[\left(\boldsymbol{\tilde{\Sigma}}^{-1} \boldsymbol{\tilde{\Sigma}}'\right)^2]}{2 \left(1+P^2\right)} = 
&\frac{2 \sinh^2 s}{(\gamma \cosh 2s -\gamma -1)(2 \gamma \cosh 2s -2 \gamma -1)} \nonumber \\
&\times \big\{\left[(1+2\gamma) \cosh 2s - 2 \gamma \right] (\partial_{\alpha_a}\eta_R)^2 + 2 \eta^2 (1+\gamma + \cosh 2s - \gamma \cosh 4s) (\partial_{\alpha_A} \theta_R)^2\big\}\,.
\label{eq:term1_QFI_gaussian}    
\end{align}
\noindent The second term corresponding to the evolution of the purity with $\alpha_A$ takes the following form
\begin{equation}
    \frac{2 {P'}^2}{1-P^4} = \frac{(1-2 \eta)^2 \sinh^2 s}{\gamma (1+\gamma-\gamma \cosh 2s)(2 \gamma \cosh 2s + 2 \gamma -1)}(\partial_{\alpha_A} \eta_R)^2\,,
\label{eq:term2_QFI_gaussian}      
\end{equation}
\noindent where $\gamma = \eta (\eta -1)$. Finally, the third term is given by
\begin{equation}
{\boldsymbol{\Delta \text{X}}'}^\intercal\, \boldsymbol{\tilde{\Sigma}}^{-1}\,\boldsymbol{\Delta \text{X}}' = \frac{\kappa_1 (\kappa_2 + \kappa_3)}{{\tau}^2 (\kappa_4 + \kappa_5 + \kappa_6)}\,,
\label{eq:term3_QFI_gaussian} 
\end{equation}
\noindent where
\begin{align*}
\kappa_1 =  &(\abs{\beta} a L \Gamma)^2 (r^2-1)^2 \,, \\
\kappa_2 =  &(a^2-1) (r^2-1) \tau  + \tau  (a^2+ r^2-2 a r \cos \phi) \cosh 2s \,, \\
\kappa_3 =  &\sinh 2 s \big\{\left[a^2 r^2 (r^2+4)+a^2\right] \cos \chi+a^3 r \left[a r \cos (2\phi -\chi )-2 (r^2+1) \cos (\phi -\chi )\right] \nonumber 
\\  &-2 a (r^3+r) \cos (\chi +\phi )+r^2 \cos (\chi +2 \phi )\big\} \,, \\
\kappa_4 =  &a^4 \left(r^4-2 r^2+2\right)+2 \left(a^2-1\right) (a^2+ r^2-2 a r \cos \phi)
  \left(r^2-1\right) \cosh (2 s) \,,\\
\kappa_5 =  &2 a r \left[a r \cos 2 \phi-2 \left(a^2+r^2\right) \cos \phi \right] \,, \\
\kappa_6 = &-2 a^2 (r^4-4 r^2+1)+2 r^4-2 r^2+1\,, \\
\tau = & 1+a^2 r^2-2 a r \cos \phi \,.
\end{align*}

We now seek to identify the parameter combinations that maximize the QFI. Due to the complex nature of the above equations which contain eight parameters, we resorted to Mathematica’s standard numerical maximization algorithm~\cite{Mathematica}. The only constraint placed on the parameter space was limiting $a \leq 0.99$, which contemplates all relevant experimental situations. The numerical algorithm converges and identifies the maximum QFI as being obtained on-resonance and at critical coupling, analogous to the coherent-state probe. Under these circumstances, the QFI reduces to \cref{eq:QFI_Gaussian} in the main text, \textit{i.e.}
\begin{equation}
    \mathcal{Q}_S = \left(\abs{\beta}^2 + \sinh^2s\right)\frac{L^2\,\Gamma^2\, B\,}{1- e^{-\,\alpha_T L }}\,.
\end{equation}

For a coherent-state probe, only the third term ${\boldsymbol{\Delta \text{X}}'}^\intercal\, \boldsymbol{\tilde{\Sigma}}^{-1}\,\boldsymbol{\Delta \text{X}}'$ in \cref{Eq:QFI_Gaussian_State} contributes. By setting $\{s, \chi\} \rightarrow 0$ in \cref{eq:term3_QFI_gaussian}, we readily obtain the corresponding QFI
\begin{equation}
Q_C = \abs{\beta}^2 L^2 \Gamma ^2 \frac{\left(1-r^2\right)^2 a^2 }{\left(1 -2 a r \cos \phi +a^2 r^2\right)^2}\,.
\end{equation}
\noindent This expression can be conveniently rewritten in terms of the buildup factor $B$ given in \cref{eq:buildup_factor}, yielding \cref{eq:QFI_coherent} in the main text $\mathcal{Q}_C= \left( \abs{\beta} \, L\, \Gamma\,B\, e^{\alpha_T L/2}\right)^2$. At critical coupling ($r=a$) and on resonance ($\phi = 2 \pi m$), $\mathcal{Q}_C$ is maximum and reduces to $\abs{\beta}^2 L^2 \Gamma^2 a^2 (a^2-1)^{-2}$. Rewriting the latter in terms of $B$ yields \cref{eq:QFI_coherent_critical} in the main text, i.e.
\begin{equation}
\mathcal{Q}_C \big\rvert_{r=a,\,\phi=2 \pi m} =\abs{\beta}^2\,\frac{L^2\,\Gamma^2\, B\,}{1- e^{-\,\alpha_T L }}\,.
\end{equation}
\noindent Replacing $\partial\alpha_A$ by $\partial n_A$ when computing the quantities in \cref{Eq:QFI_Gaussian_State}, an analogous procedure can be followed to derive the QFI in estimating the analyte refractive index $n_A$ given in \cref{eq:QFI_Quantum_Limit_nA} in the main text.

\end{document}